\documentclass[preprint,authoryear,12pt]{elsarticle}

\usepackage{amssymb}

\journal{ }

\begin{document}

\begin{frontmatter}


\author{Michal Brzezinski}
\ead{mbrzezinski@wne.uw.edu.pl}
\address{Faculty of Economic Sciences, University of Warsaw, Poland}

\title{Empirical modeling of the impact factor distribution}


\author{}

\address{}

\begin{abstract}
The distribution of impact factors has been modeled in the recent informetric literature using two-exponent law proposed by \cite{mansilla2007behavior}. This paper shows that two distributions widely-used in economics, namely the Dagum and Singh-Maddala models, possess several advantages over the two-exponent model. Compared to the latter, the former give as good as or slightly better fit to data on impact factors in eight important scientific fields. In contrast to the two-exponent model, both proposed distributions have closed-from probability density functions and cumulative distribution functions, which facilitates fitting these distributions to data and deriving their statistical properties.

\end{abstract}

\begin{keyword}
impact factor \sep two-exponent law \sep Dagum model \sep Singh-Maddala model \sep maximum likelihood estimation  \sep
 model selection 



\end{keyword}

\end{frontmatter}


\section{Introduction}
\label{intro}

The distribution of journal impact factors has been recently studied in the informetric literature from both theoretical and empirical perspectives. 
\cite{mansilla2007behavior}
proposed the two-exponent law to model rank-frequency distributions of impact factors. 
This model was used by \cite{campanario2010a} to study empirically changes in the distribution of impact factors over time. A theoretical derivation of the rank-frequency distribution of impact factors was derived by \cite{egghe2009mathematical}; see also \cite{egghe2011impact} and \cite{egghe2011relations}.\footnote{\cite{waltman2009some} have criticized the approach of \cite{egghe2009mathematical} from an empirical point of view.} 
\cite{mishra2010note} has fitted several well-established statistical distributions to data on impact factors for journals from several scientific disciplines. The two-exponent law introduced by \cite{mansilla2007behavior} has been recently studied by \cite{sarabia2012modeling}. The authors have obtained the probabilistic quantile function corresponding to the two-exponent law as well as derived several statistical measures and tools associated with this law like the moments, Lorenz and Leimkuhler curves and the Gini index of inequality. Moreover, they fitted the two-exponent law to data on impact factors for eight science categories and found that the fit of the model was satisfactory.

The present paper contributes to the literature on empirical modeling of the journal impact factor distribution by verifying in a statistically rigorous way if the two-exponent law is consistent with data on impact factors. In particular, the paper fits the two-exponent law do data using the maximum likelihood approach, which is more efficient than the least squares approach used in previous studies \citep{mansilla2007behavior,sarabia2012modeling}. The fit of the model to data is evaluated using an appropriate goodness-of-fit test. Finally, the two-exponent law is compared to alternative models using a likelihood ratio test. As alternatives we have chosen the Singh-Maddala and Dagum statistical models \citep[see, e.g.,][]{kleiber2003statistical}, which are widely-used in economics to model the distribution of income and other variables.\footnote{These distributions, among others, were previously used by \cite{mishra2010note} to study the distribution of impact factors. However, they were not compared with the two-exponent model.}
 The analysis is performed for impact factors from eight science categories studied previously by \cite{sarabia2012modeling}.

The reminder of the paper is structured as follows. Section \ref{sec:methods} introduces definition and basic properties of the two-exponent law and the compared alternative models. Statistical tests used to assess goodness of fit to data and tests used for model selection are presented as well. Section \ref{sec:data} describes our data on impact factors, while Section \ref{sec:results} presents and discusses empirical findings. Finally, Section \ref{sec:conclusions} concludes.

\section{Methods}
\label{sec:methods}

\subsection{The two-exponent law versus Dagum and Singh-Maddala models}

In order to model the distribution of impact factors, \cite{mansilla2007behavior} introduced the two-exponent law in terms of rank-frequency distribution taking the form:
\begin{equation}
f(r)=K\frac{(N+1-r)^b}{r^a},
\label{eq:1}
\end{equation}
where $a>0$, $b>0$ are shape parameters, $K>0$ is a scale parameter, $N$ is the total number of sources (journals in the case of modeling impact factors), $r=1,2,..., N$ is the ranking number and $f(r)$ it the impact factor. If $b=0$, then (\ref{eq:1}) reduces to Zipf's law \citep{egghe2005power}; if $a=b$ it becomes the Lavalette law \citep{lavalette1996facteur}, and when $a=0$ it becomes a power law. \cite{sarabia2012modeling} derive the quantile function that corresponds to the two-exponent law (\ref{eq:1}), which takes the form:
\begin{equation}
F^{-1}(u)= K\frac{u^b}{(1-u)^a}, \; \mathrm{for} \; 0 < u < 1.
\label{eq:2}
\end{equation}
The distribution defined by (\ref{eq:2}) was introduced in statistical literature by \cite{hankin2006new}, who called it the Davies distribution.
Unfortunately, neither probability density function (pdf) nor cumulative distribution function (cdf) for (\ref{eq:2}) is available in closed form except for some special cases like the Zipf's law, the power law or the Lavalette law. However, they can be calculated numerically by inverting the quantile function (\ref{eq:2}).

The Singh-Maddala and Dagum models were introduced in economics in the context of modeling income distribution by, respectively, \cite{singhmaddala} and \cite{dagum1977}. In statistics, these distributions appeared first in the system of distributions of \cite{burr1942cumulative} and are known as Burr XII distribution (Singh-Maddala) and Burr III distribution (Dagum). For the Singh-Maddala and Dagum distributions, the pdfs, cdfs and quantile functions
are available in closed forms. Specifically,  for a sample of positive impact factors in a given scientific filed, $x_1, ..., x_N$, the cdf for the Singh-Maddala distribution is given by:
\begin{equation}
F(x)=1-\left[ 1+\left( \frac{x}{b} \right)^a \right]^{-q}, 
\end{equation} where $a>0$, $q>0$ are shape parameters and $b>0$ is a scale parameter. The cdf for the Dagum distribution is
\begin{equation}
F(x)=\left[ 1+\left( \frac{x}{b} \right)^{-a} \right]^{-p}, 
\end{equation}
where $a>0$, $p>0$ are shape parameters and $b>0$ is a scale parameter. The Singh-Maddala and Dagum distributions are closely related in the following way: \(X\sim \mathrm{Dagum}(a,b,p) \Longleftrightarrow\frac{1}{X} \mathrm{SM}(a,\frac{1}{b},p))\), where $\sim$ means ``is distributed as".
The upper tail of the Singh-Maddala distribution is governed by two parameters ($a$ and $q$), while the lower tail by $a$ only \citep{kleiber1996dagum}.
The opposite holds for the Dagum model (two parameters govern the behaviour of the lower tail and only one shapes the upper tail) and for this reason this model is more flexible in the lower tail. Therefore, the two models can be considered complementary as they have advantages in modeling different parts of the data. Theoretical properties of the Singh-Maddala and Dagum distributions are very well known; see \cite{kleiber1996dagum,kleiber2008dagum} and \cite{kleiber2003statistical} for a detailed discussion.\footnote{The Singh-Maddala and Dagum distributions are nested within a four-parameter Generalized Beta of the Second Kind (GB2) model introduced by \cite{mcdonald1984some}. We have experimented with fitting this model to data on impact factors, but the gains from the additional complexity were small.}
In particular, while \cite{sarabia2012modeling} offer expressions for the Lorenz curve and the Gini index of inequality for the two-exponent distribution, in case of the Singh-Maddala and Dagum models expressions for a wide variety of inequality measures exist \citep{kleiber2003statistical,jenkins2009distributionally}. Moreover, the conditions that allow for testing Lorenz dominance (i.e. inference on inequality robust to the choice of an inequality measure) are available \citep{wilfling1993lorenz,kleiber1996dagum}.

When $a=b$ the two-exponent law becomes the Lavalette law \citep{lavalette1996facteur}. As noticed by \cite{sarabia2012modeling}, the Lavalette law is known in the economic literature as the Fisk (or log-logistic) distribution. The Fisk distribution is a special case of both the Singh-Maddala model (with $q$ set to 1) and the Dagum model (with $p$ set to 1). \cite{kleiber2003statistical} discuss the properties of the Fisk distribution in more detail.

The Singh-Maddala and Dagum distributions were shown to fit income distributions well. \cite{mcdonald1984some} studied the fit of several three- and four-parameter statistical models to grouped income distribution data and found that the Singh-Maddala model performed best among the three-parameter distributions and even better than one four-parameter distribution. 
\cite{dastrup2007impact} found that for the disposable income variable the Singh-Maddala model was the best-fitting distribution as often as the Dagum distribution in the group of three-parameter distributions.
On the other hand, \cite{bandourian2003comparison} compared several distributional models for 82 data sets on gross income data and found that the Dagum model is the best three-parameter model in 84\% of cases.

\subsection{Estimation methods, goodness-of-fit and model selection tests}
\label{sec:estimation}
We estimate all models using maximum likelihood (ML) estimation. The log-likelihood functions for the Singh-Maddala and\ Dagum models are available in \cite{kleiber2003statistical}. The log-likelihood for the two-exponent model can be calculated by numerical inversion of the quantile function (\ref{eq:2}) \citep{hankin2006new}.\ Parameter estimates are obtained by numerical maximization of the log-likelihood functions. The variance matrix is calculated as the negative inverse of the Hessian
evaluated at the parameter estimates.

\cite{sarabia2012modeling} estimate the parameters of the two-exponent law using ordinary least squares\ (OLS)\ approach, which was earlier proposed by \cite{hankin2006new}. However, \cite{hankin2006new} show in a simulation study that the OLS approach should be used only when the exponents $a$ and $b$ in (\ref{eq:2}) are equal. Since there is no theoretical reason to assume that, this paper uses rather the ML approach to estimate the parameters and their variances of the two-exponent law.\footnote{Notice also that \cite{sarabia2012modeling} estimate standard errors for parameter estimates of the two-exponent model by directly taking the standard errors from the OLS regression of logged sample order statistics (or observed quantiles) on their expected values. However, this approach does not take into account the covariance between order statistics and produces erroneous results. \cite{hankin2006new} provide a correct method of calculating standard errors for parameters of (\ref{eq:2}) estimated using the OLS.}          

In order to assess whether our samples of impact factors are consistent with a given statistical model, we follow \cite{clauset2009power} in using a goodness-of-fit test based on parametric bootstrap approach. In particular, we use the well-known Kolmogorov-Smirnov test  defined by:
\begin{equation}
KS = \max\limits_{x} \left\vert F(x) - F(x; \hat\theta)\right\vert,
\end{equation}
where $F(x)$ is the cdf of the data, $F(x; \hat\theta)$ is the cdf of a tested model and $\hat\theta$ is the vector of the model's parameter estimates obtained, for example, using ML estimation. The distribution of the $KS$ statistic is known for data sets drawn from a given model. However, when the underlying model is not known or when its parameters are estimated from the data, which is the case studied in this paper, the distribution of the $KS$ must be obtained by simulation. The appropriate simulation is implemented in the following way. First, we fit a given model, $F(x; \theta)$, to our data set obtaining a vector of estimated parameters $\hat\theta$. Next, we calculate a $KS$ statistic for the fitted model and denote it by $KS_{org}$. Third, we draw a large number, $B$, of synthetic samples of the original size from a fitted model $F(x; \hat\theta)$. For each simulated sample, we fit the model $F(x; \theta)$ and calculate its $KS$ statistic denoted by $KS_b$. Finally, the $p$-value for the test is obtained as the proportion of $KS_b$ greater than $KS_{org}$. The hypothesis that our data set follows $F(x; \theta)$ is rejected if the $p$-value is smaller than the chosen threshold (set to 0.1 in this paper). 

We also compare formally whether the two-exponent law gives a better fit to the impact factor data than the the Singh-Maddala and Dagum distributions. To this aim, we use the likelihood ratio test, which tests if the compared models are equally close to the true model against the alternative that one is closer.
The test computes the logarithm of the ratio of the likelihoods of the data under two competing distributions, $LR$, which is negative or positive depending on which model fits data better. \cite{vuong1989likelihood} showed that in the case of non-nested models the normalized
log-likelihood ratio $NLR=n^{-1/2}LR/\sigma$, where $\sigma$ is the estimated standard deviation of $LR$, has a limit standard normal distribution. This result can be used to compute a $p$-value for the test discriminating between the competing models.

\section{Data}
\label{sec:data}

We use data on impact factors from the latest available (2012) edition of Thompson Reuters Journal Citation Reports (JCR). Following
\cite{sarabia2012modeling}, we use impact factors for scientific journals belonging to the following scientific fields: Chemistry, Economics, Education, Information Science and Library Science (abbreviated further as Information SLS), Mathematics, Neurosciences, Psychology and Physics.\footnote{In some cases, we have grouped several JCR subject categories into one science category. For example, our category ``Mathematics'' consists of the following JCR subject categories: ``Mathematical \& Computational Biology'', ``Mathematics'', ``Mathematics, Applied'' and ``Mathematics, Interdisciplinary Applications''.}
We have removed journals with zero impact factors from our samples. Descriptive statistics for our data sets are presented in Table \ref{table:table1}. 

\begin{table}[h]
\centering
\footnotesize
\caption{Descriptive statistics for the impact factors in eight scientific fields ($N$ denotes the number of journals within a given field).}
\label{table:table1}
\begin{tabular}{lccccc}
\hline 
Field   & $N$ & Mean & Median & Std. Dev. & Gini   \\ \hline
Chemistry & 513 & 2.703 & 1.684 & 3.669 & 0.513 \\
Economics & 333 & 1.062 & 0.795 & 0.922 & 0.440 \\
Education & 254 & 0.868 & 0.679 & 0.686 & 0.404 \\
Information SLS & 84 & 1.001 & 0.755 & 0.928 & 0.465 \\
Mathematics & 573 & 0.950 & 0.717 & 0.769 & 0.379 \\
Neurosciences & 252 & 3.574 & 2.872 & 3.486 & 0.419 \\
Psychology & 556 & 1.766 & 1.361 & 1.852 & 0.435 \\
Physics & 381 & 2.535 & 1.400 & 4.514 & 0.568 \\ \hline
\end{tabular}
\end{table}

Comparing our data with data used by \cite{sarabia2012modeling}, we observe that the number of journals indexed in JCR has increased between JCR 2010 Edition and JCR 2012 edition for each science category analyzed; the increases lie in the range between 5\% and 20\%. Other descriptive statistics are close to those reported by \cite{sarabia2012modeling}.

\section{Empirical results and discussion}
\label{sec:results}

Table \ref{table:twoexp} presents results of fitting the two-exponent law to the data. Beside parameter estimates and their standard errors, we give also the value of log-likelihood and the $p$-value from the goodness-of-fit test described in Section \ref{sec:estimation}. 

\begin{table}[h]
\centering
\footnotesize
\caption{Maximum likelihood fits of the two-exponent law to data on impact factors in selected fields of science. Standard errors in parentheses.} 
\label{table:twoexp}
\begin{tabular}{lccccc} 
\hline
Field   & $K$ & $b$ & $a$ &  Log-likelihood & $p$-value \\ \hline
Chemistry & 1.8972 (0.1371) & 0.6331 (0.0459) & 0.4997 (0.0372) & -982.63 & 0.650 \\
Economics & 1.1403 (0.0892) & 0.7827 (0.0618) & 0.3197 (0.0361) & -343.19 & 0.615\\
Education & 0.8453 (0.0791) & 0.6155 (0.0615) & 0.3334 (0.0448) & -198.02 & 0.013\\
Information SLS & 1.0459 (0.1832) & 0.8308 (0.1246) & 0.3466 (0.0841) & -82.99 & 0.006\\
Mathematics & 0.6730 (0.0377) & 0.3334 (0.0290) & 0.4386 (0.0337) & -427.29 & 0.105\\
Neurosciences & 3.4131 (0.2438) & 0.6386 (0.0558) & 0.3213 (0.0330) & -549.75 &  0.002\\
Psychology & 1.5840 (0.0845) & 0.6096 (0.0384) & 0.3577 (0.0278) & -819.86 & 0.314 \\
Physics & 1.2452 (0.0827) & 0.4722 (0.0388) & 0.5965 (0.0444) & -657.62 & 0.008 \\ \hline
\end{tabular}
\end{table}

Our estimates of the parameters are in general close to the OLS estimates obtained by \cite{sarabia2012modeling}. However, our estimates of standard errors are roughly one order higher than those of \cite{sarabia2012modeling}. As pointed out before, this is due to the fact that \cite{sarabia2012modeling} do not account in their approach for the dependence between order statistics. For this reason, their standard errors are severely underestimated. 

The goodness-of-fit test used suggests that the two-exponent model is a plausible hypothesis for impact factors of journals in Chemistry, Economics, Mathematics and Psychology.

We can use our estimates to test for the Lavalette law, which holds that $a=b$. Using Wald test \citep[see][]{sarabia2012modeling}, we cannot reject the hypothesis that $a=b$ at the 5\% level for Mathematics and Physics (with $p$-values equal to 0.063 and 0.075, respectively). Also, the significance level of the test is close to 5\% for Chemistry ($p$-value = 0.048).
For other science categories, the hypothesis tested is rejected even at the 1\% level. Our results with respect to the Lavalette law are inconsistent with those of \cite{sarabia2012modeling}, who found that the law was rejected for each of the science category studied. This inconsistency is due to the previously mentioned fact that \cite{sarabia2012modeling} substantially underestimated the variance of the parameter estimates of the two-exponent law.

Tables \ref{table:sm}-\ref{table:dagum} presents our results of fitting the Singh-Maddala and Dagum models. In most of the cases, the parameters are estimated with sufficient precision. There are some exceptions, especially in case of fitting the Singh-Maddala distribution to impact factors for Economics, Education, and Information SLS. However, even in these cases all three parameters are jointly different from zero in a statistically significant way.

\begin{table}[h]
\centering
\footnotesize
\caption{Maximum likelihood fits of the Singh-Maddala model to data on impact factors in selected fields of science. Standard errors in parentheses.} 
\label{table:sm}
\begin{tabular}{lccccc} 
\hline
Field   & $a$ & $b$ & $q$ &  Log-likelihood & $p$-value \\ \hline
Chemistry & 1.6208 (0.1030) & 2.2040 (0.3789) & 1.3613 (0.2477) & -982.60 & 0.759 \\
Economics & 1.3534 (0.1004) & 3.6112 (2.1859) & 5.5194 (3.4980) & -341.64 & 0.916 \\
Education & 1.6679 (0.1536) & 1.4947 (0.5821) & 2.8249 (1.2977) & -197.07 & 0.061 \\
Information SLS & 1.2895 (0.1962) & 3.1271 (3.8256) & 4.7948 (5.7605) & -82.44 & 0.043 \\
Mathematics & 2.8287 (0.2140) & 0.6471 (0.0602) & 0.7957 (0.1303) & -428.15 & 0.037 \\
Neurosciences & 1.7098 (0.1354) & 4.8304 (1.1260) & 2.1769 (0.5982) & -552.07 & 0.001 \\
Psychology & 1.7595 (0.0980) & 2.0466 (0.3225) & 1.8321 (0.3432) & -821.01 & 0.085 \\
Physics & 2.1123 (0.1623) & 1.0978 (0.1274) & 0.7382 (0.1133) & -657.02 & 0.038 \\ \hline
\end{tabular}
\end{table}

\begin{table}[h]
\centering

\footnotesize
\caption{Maximum likelihood fits of the Dagum model to data on impact factors in selected fields of science. Standard errors in parentheses.} 
\label{table:dagum}
\begin{tabular}{lccccc} 
\hline
Field   & $a$ & $b$ & $p$ &  Log-likelihood & $p$-value \\ \hline
Chemistry & 1.9727 (0.1454) & 2.0778 (0.2616) & 0.7701 (0.1191) & -982.87 & 0.528 \\
Economics & 2.7042 (0.3041) & 1.3517 (0.1585) & 0.4420 (0.0835) & -344.56 & 0.648 \\
Education & 2.7038 (0.3170) & 0.9520 (0.1276) & 0.5702 (0.1248) & -198.64 & 0.016 \\
Information SLS & 2.4773 (0.7232) & 1.2474 (0.4136) & 0.4605 (0.2290) & -83.73 & 0.005 \\
Mathematics & 2.3283 (0.1400) & 0.5924 (0.0704) & 1.4299 (0.2592) & -426.87 & 0.146 \\
Neurosciences & 3.0261 (0.3013) & 4.1831 (0.3626) & 0.4615 (0.0741) & -547.63 & 0.014\\
Psychology & 2.6482 (0.1986) & 1.8506 (0.1547) & 0.5701 (0.0775) & -819.85 & 0.293 \\
Physics & 1.7539 (0.1255) & 1.1489 (0.1890) & 1.2473 (0.2334) & -658.09 & 0.020 \\ \hline
\end{tabular}
\end{table}

The goodness-of-fit test suggests that both models are plausible for impact factors of journals in Chemistry and Economics. The Dagum model seems also to be consistent with data for Mathematics and Psychology. The Singh-Maddala model has a higher probability of being consistent with data for Education, Information SLS, and Physics than both the two-exponent law and the Dagum model. Neither model seems to be consistent with data for Neurosciences.

We can also test for the Lavalette law (Fisk distribution) using fitted Singh-Maddala and Dagum models. For the Singh-Maddala model this amounts to testing the hypothesis that $q=1$, while for the Dagum model the relevant hypothesis is that $p=1$. Using Wald tests, we have found that at the 5\% level both the Singh-Maddala and Dagum models can be reduced to the Fisk distribution in case of Chemistry and Mathematics. Moreover, the Singh-Maddala model can be reduced to the Fisk distribution in case of Economics, Education, and Information SLS, while the Dagum model reduces to the Fisk model in case of Physics.\footnote{However, these results for the Singh-Maddala model may be due to the previously discussed insufficient precision of the estimates for the parameter $q$.} These results suggest that in modeling the distribution of impact factors, the Lavalette law should be treated as a serious alternative to the three-parameter distributions, especially for formal and natural sciences.

The results of model selection tests based on the approach of \cite{vuong1989likelihood} are reported in Table \ref{table:modelsel}. The compared models are empirically indistinguishable for impact factors of journals in Chemistry, Mathematics, Psychology and Physics. For Economics, Education, and Information SLS, the Singh-Maddala model is preferred to both the two-exponent model and the Dagum model. Finally, the Dagum distribution is preferred for data in Neurosciences.

\begin{table}[h]
\centering
\footnotesize
\caption{Model selection tests for impact factor distributions, selected science fields. $``\mathrm{A} \succ \mathrm{B}"$ denotes that distribution A gives a better fit than distribution B. Positive values of $LR$ indicate that the likelihood for the two-exponent model is higher than the likelihood for the alternative.}
\label{table:modelsel}
\begin{tabular}{lccccc}  \hline
Science category &  \multicolumn{2}{c}{Two-exp. vs. Dagum} & \multicolumn{2}{c}{Two-exp. vs. SM} &  Conclusion  \\  
 & $LR$ & \textit{p}-value & $LR$ & \textit{p}-value &  \\  \hline
Chemistry & 0.237 & 0.483 & -0.027 & 0.939 & Distributions are indistinguishable \\
Economics & 1.365 & 0.071 & -1.554 & 0.094 & Two-exp. $\succ$ Dagum, SM $\succ$ Two-exp.  \\
Education & 0.623 & 0.184 & -0.949 & 0.058 & Two-exp. $\succ$ Dagum, SM $\succ$ Two-exp.  \\
Information SLS & 0.746 & 0.022 & -0.545 & 0.089 & Two-exp. $\succ$ Dagum, SM $\succ$ Two-exp. \\
Mathematics & -0.416 & 0.336 & 0.864 & 0.117 & Distributions are indistinguishable \\
Neurosciences & -2.115 & 0.005 & 2.322 & 0.020 & Dagum $\succ$ Two-exp., Two-exp. $\succ$ SM  \\
Psychology & -0.017 & 0.981 & 1.148 & 0.214 & Distributions are indistinguishable \\
Physics & 0.472 & 0.234 & -0.599 & 0.137 & Distributions are indistinguishable \\ \hline

\end{tabular}
\end{table}

\begin{figure}[!htbp]
\centering
\includegraphics[scale=0.5]{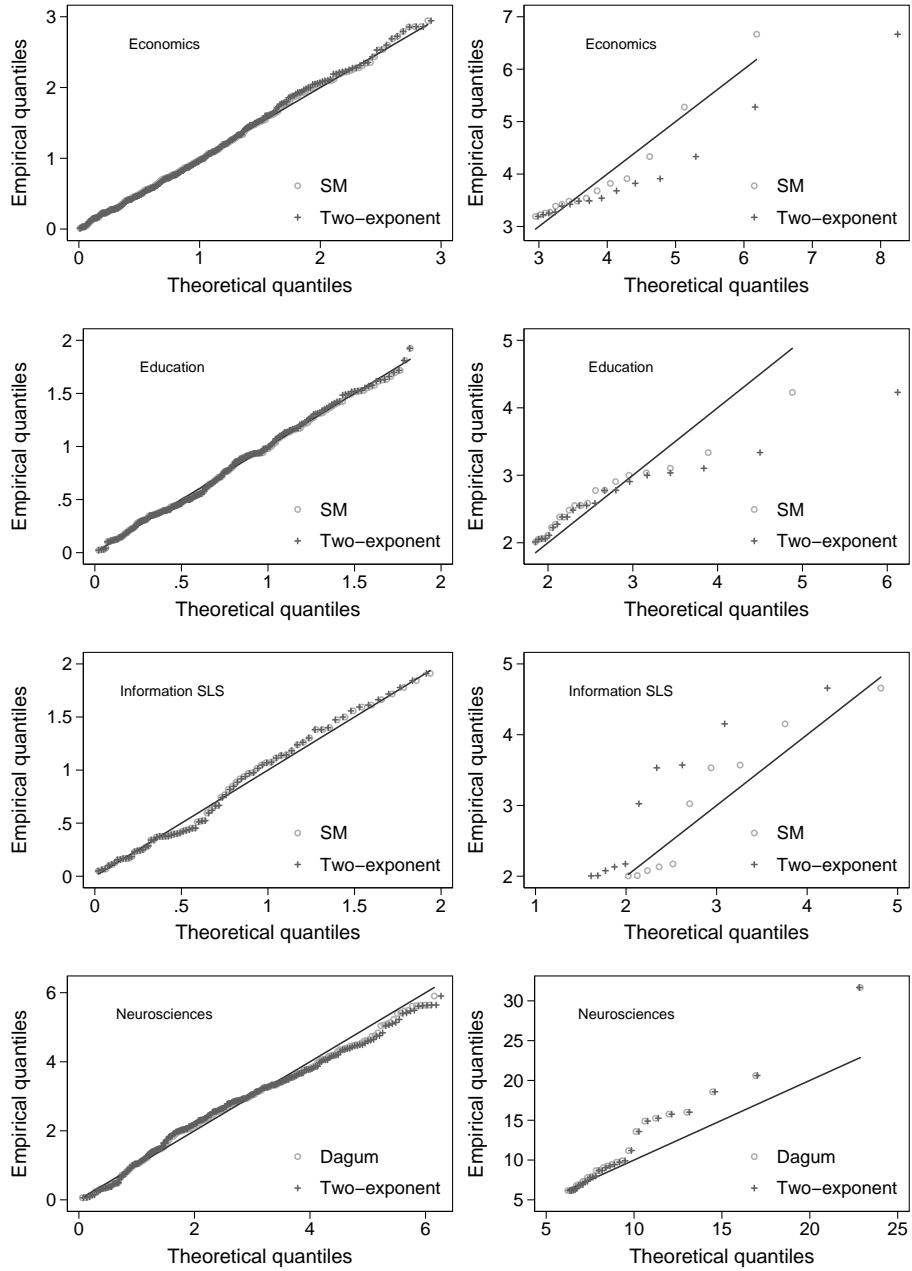}
\caption{Quantile-quantile plots for the two-exponent law vs. the better of the Singh-Maddala and Dagum models. Left panels show bottom and middle quantiles, right panels show upper quantiles.}
\label{figure1}
\end{figure}

The models can be also compared visually using quantile-quantile (q-q) plots. Figure \ref{figure1} plots empirical quantiles versus theoretical quantiles for the two-exponent law and the better of the two alternatives (Singh-Maddala or Dagum) for these cases, when results from Table \ref{table:modelsel} suggest that the two-exponent law is worse fit to the data. If the estimated model fitted the data perfectly, the q-q plot would coincide with the 45-degree line. Figure \ref{figure1} shows that the differences between compared models are in general small. However, it is clearly visible that the two-exponent law gives a slightly worse fit in these cases, especially for the highest quantiles. 

Overall, both our formal tests and visual inspection from the q-q plots suggest that compared to the two-exponent law, the Singh-Maddala and Dagum models offer as good as or slightly better fit to data on impact factors. 

\section{Conclusions}
\label{sec:conclusions}

The distribution of impact factors has been modeled in the recent informetric
literature using the two-exponent law proposed by Mansilla et al. (2007).
This paper shows that two distributions widely-used in economics, namely
the Singh-Maddala and Dagum models, possess several advantages over the two-exponent
model. They are either empirically indistinguishable from the two-exponent
law or give slightly better fit to data on impact factors in eight important
scientific fields. Contrary to \cite{sarabia2012modeling}, the present paper also found that the Lavalette law \citep{lavalette1996facteur}, which is a special case of both the two-exponent law and the Singh-Maddala and Dagum models, often cannot be rejected as an appropriate distribution for impact factors, especially for formal and natural sciences.

In contrast to the two-exponent model, both models proposed in this paper
have closed-from probability density functions and cumulative distribution functions. This advantage allows one to estimate model parameters in a more straightforward way without the need for numerical estimation of the model's basic functions. There exists also a well-developed literature \citep[see, e.g.,][]{kleiber2003statistical,kleiber2008dagum},  that explores various theoretical properties of the Singh-Maddala and Dagum distributions, including ready-made expressions for several inequality measures as well as conditions required for Lorenz ordering or stochastic dominance of various degrees.
These properties may be exploited in modeling the distribution of impact factors and other informetric variables. For these reasons, this paper suggests that the Singh-Maddala and Dagum models are more useful for empirical modeling of the distribution of journal impact factors than the two-exponent law.

\bibliographystyle{model5-names}
\bibliography{if_modeling}

\begin{thebibliography}{24}
\expandafter\ifx\csname natexlab\endcsname\relax\def\natexlab#1{#1}\fi
\providecommand{\url}[1]{\texttt{#1}}
\providecommand{\href}[2]{#2}
\providecommand{\path}[1]{#1}
\providecommand{\DOIprefix}{doi:}
\providecommand{\ArXivprefix}{arXiv:}
\providecommand{\URLprefix}{URL: }
\providecommand{\Pubmedprefix}{pmid:}
\providecommand{\doi}[1]{\href{http://dx.doi.org/#1}{\path{#1}}}
\providecommand{\Pubmed}[1]{\href{pmid:#1}{\path{#1}}}
\providecommand{\bibinfo}[2]{#2}
\ifx\xfnm\relax \def\xfnm[#1]{\unskip,\space#1}\fi
\bibitem[{Bandourian et~al.(2003)Bandourian, McDonald \&
  Turley}]{bandourian2003comparison}
\bibinfo{author}{Bandourian, R.}, \bibinfo{author}{McDonald, J.~B.}, \&
  \bibinfo{author}{Turley, R.~S.} (\bibinfo{year}{2003}).
\newblock \bibinfo{title}{A comparison of parametric models of income
  distribution across countries and over time}.
\newblock {\it \bibinfo{journal}{Estad\'{i}stica}\/},  {\it
  \bibinfo{volume}{55}\/}, \bibinfo{pages}{135--152}.
\bibitem[{Burr(1942)}]{burr1942cumulative}
\bibinfo{author}{Burr, I.~W.} (\bibinfo{year}{1942}).
\newblock \bibinfo{title}{Cumulative frequency functions}.
\newblock {\it \bibinfo{journal}{The Annals of Mathematical Statistics}\/},
  {\it \bibinfo{volume}{13}\/}, \bibinfo{pages}{215--232}.
\bibitem[{Campanario(2010)}]{campanario2010a}
\bibinfo{author}{Campanario, J.~M.} (\bibinfo{year}{2010}).
\newblock \bibinfo{title}{Distribution of changes in impact factors over time}.
\newblock {\it \bibinfo{journal}{Scientometrics}\/},  {\it
  \bibinfo{volume}{84}\/}, \bibinfo{pages}{35--42}.
\bibitem[{Clauset et~al.(2009)Clauset, Shalizi \& Newman}]{clauset2009power}
\bibinfo{author}{Clauset, A.}, \bibinfo{author}{Shalizi, C.~R.}, \&
  \bibinfo{author}{Newman, M.~E.} (\bibinfo{year}{2009}).
\newblock \bibinfo{title}{Power-law distributions in empirical data}.
\newblock {\it \bibinfo{journal}{SIAM review}\/},  {\it
  \bibinfo{volume}{51}\/}, \bibinfo{pages}{661--703}.
\bibitem[{Dagum(1977)}]{dagum1977}
\bibinfo{author}{Dagum, C.} (\bibinfo{year}{1977}).
\newblock \bibinfo{title}{A new model of personal income distribution:
  Specification and estimation}.
\newblock {\it \bibinfo{journal}{Economie Appliqu{\'e}e}\/},  {\it
  \bibinfo{volume}{30}\/}, \bibinfo{pages}{413--437}.
\bibitem[{Dastrup et~al.(2007)Dastrup, Hartshorn \&
  McDonald}]{dastrup2007impact}
\bibinfo{author}{Dastrup, S.~R.}, \bibinfo{author}{Hartshorn, R.}, \&
  \bibinfo{author}{McDonald, J.~B.} (\bibinfo{year}{2007}).
\newblock \bibinfo{title}{The impact of taxes and transfer payments on the
  distribution of income: A parametric comparison}.
\newblock {\it \bibinfo{journal}{The Journal of Economic Inequality}\/},  {\it
  \bibinfo{volume}{5}\/}, \bibinfo{pages}{353--369}.
\bibitem[{Egghe(2005)}]{egghe2005power}
\bibinfo{author}{Egghe, L.} (\bibinfo{year}{2005}).
\newblock {\it \bibinfo{title}{Power laws in the information production
  process: Lotkaian informetrics}\/}.
\newblock \bibinfo{address}{Oxford}: \bibinfo{publisher}{Elsevier}.
\bibitem[{Egghe(2009)}]{egghe2009mathematical}
\bibinfo{author}{Egghe, L.} (\bibinfo{year}{2009}).
\newblock \bibinfo{title}{Mathematical derivation of the impact factor
  distribution}.
\newblock {\it \bibinfo{journal}{Journal of Informetrics}\/},  {\it
  \bibinfo{volume}{3}\/}, \bibinfo{pages}{290--295}.
\bibitem[{Egghe(2011)}]{egghe2011impact}
\bibinfo{author}{Egghe, L.} (\bibinfo{year}{2011}).
\newblock \bibinfo{title}{The impact factor rank-order distribution revisited}.
\newblock {\it \bibinfo{journal}{Scientometrics}\/},  {\it
  \bibinfo{volume}{87}\/}, \bibinfo{pages}{683--685}.
\bibitem[{Egghe \& Waltman(2011)}]{egghe2011relations}
\bibinfo{author}{Egghe, L.}, \& \bibinfo{author}{Waltman, L.}
  (\bibinfo{year}{2011}).
\newblock \bibinfo{title}{Relations between the shape of a size-frequency
  distribution and the shape of a rank-frequency distribution}.
\newblock {\it \bibinfo{journal}{Information Processing \& Management}\/},
  {\it \bibinfo{volume}{47}\/}, \bibinfo{pages}{238--245}.
\bibitem[{Hankin \& Lee(2006)}]{hankin2006new}
\bibinfo{author}{Hankin, R.~K.}, \& \bibinfo{author}{Lee, A.}
  (\bibinfo{year}{2006}).
\newblock \bibinfo{title}{A new family of non-negative distributions}.
\newblock {\it \bibinfo{journal}{Australian \& New Zealand Journal of
  Statistics}\/},  {\it \bibinfo{volume}{48}\/}, \bibinfo{pages}{67--78}.
\bibitem[{Jenkins(2009)}]{jenkins2009distributionally}
\bibinfo{author}{Jenkins, S.~P.} (\bibinfo{year}{2009}).
\newblock \bibinfo{title}{{Distributionally-sensitive inequality indices and
  the GB2 income distribution}}.
\newblock {\it \bibinfo{journal}{Review of Income and Wealth}\/},  {\it
  \bibinfo{volume}{55}\/}, \bibinfo{pages}{392--398}.
\bibitem[{Kleiber(1996)}]{kleiber1996dagum}
\bibinfo{author}{Kleiber, C.} (\bibinfo{year}{1996}).
\newblock \bibinfo{title}{{Dagum vs. Singh-Maddala income distributions}}.
\newblock {\it \bibinfo{journal}{Economics Letters}\/},  {\it
  \bibinfo{volume}{53}\/}, \bibinfo{pages}{265--268}.
\bibitem[{Kleiber(2008)}]{kleiber2008dagum}
\bibinfo{author}{Kleiber, C.} (\bibinfo{year}{2008}).
\newblock \bibinfo{title}{{A Guide to the Dagum Distribution}}.
\newblock In \bibinfo{editor}{D.~Chotikapanich} (Ed.), {\it
  \bibinfo{booktitle}{Modelling Income Distributions and Lorenz Curves: Essays
  in Memory of Camilo Dagum}\/} (pp. \bibinfo{pages}{97--268}).
\newblock \bibinfo{address}{Berlin, New York}: \bibinfo{publisher}{Springer}.
\bibitem[{Kleiber \& Kotz(2003)}]{kleiber2003statistical}
\bibinfo{author}{Kleiber, C.}, \& \bibinfo{author}{Kotz, S.}
  (\bibinfo{year}{2003}).
\newblock {\it \bibinfo{title}{Statistical size distributions in economics and
  actuarial sciences}\/}.
\newblock \bibinfo{address}{Hoboken, NJ}: \bibinfo{publisher}{John Wiley}.
\bibitem[{Lavalette(1996)}]{lavalette1996facteur}
\bibinfo{author}{Lavalette, D.} (\bibinfo{year}{1996}).
\newblock \bibinfo{title}{Facteur d'impact: impartialit{\'e} ou impuissance}.
\newblock \bibinfo{note}{Report, INSERM U350, 91405, Orsay (France): Institut
  Curie—Recherche, B{\^a}t. 112}.
\bibitem[{Mansilla et~al.(2007)Mansilla, K{\"o}ppen, Cocho \&
  Miramontes}]{mansilla2007behavior}
\bibinfo{author}{Mansilla, R.}, \bibinfo{author}{K{\"o}ppen, E.},
  \bibinfo{author}{Cocho, G.}, \& \bibinfo{author}{Miramontes, P.}
  (\bibinfo{year}{2007}).
\newblock \bibinfo{title}{On the behavior of journal impact factor rank-order
  distribution}.
\newblock {\it \bibinfo{journal}{Journal of Informetrics}\/},  {\it
  \bibinfo{volume}{1}\/}, \bibinfo{pages}{155--160}.
\bibitem[{McDonald(1984)}]{mcdonald1984some}
\bibinfo{author}{McDonald, J.~B.} (\bibinfo{year}{1984}).
\newblock \bibinfo{title}{Some generalized functions for the size distribution
  of income}.
\newblock {\it \bibinfo{journal}{Econometrica}\/},  {\it
  \bibinfo{volume}{52}\/}, \bibinfo{pages}{647--663}.
\bibitem[{Mishra(2010)}]{mishra2010note}
\bibinfo{author}{Mishra, S.~K.} (\bibinfo{year}{2010}).
\newblock \bibinfo{title}{A note on empirical sample distribution of journal
  impact factors in major discipline groups}.
\newblock \bibinfo{note}{Available at:
  http://works.bepress.com/sk{\_}mishra/8}.
\bibitem[{Sarabia et~al.(2012)Sarabia, Prieto \& Trueba}]{sarabia2012modeling}
\bibinfo{author}{Sarabia, J.~M.}, \bibinfo{author}{Prieto, F.}, \&
  \bibinfo{author}{Trueba, C.} (\bibinfo{year}{2012}).
\newblock \bibinfo{title}{Modeling the probabilistic distribution of the impact
  factor}.
\newblock {\it \bibinfo{journal}{Journal of Informetrics}\/},  {\it
  \bibinfo{volume}{6}\/}, \bibinfo{pages}{66--79}.
\bibitem[{Singh \& Maddala(1976)}]{singhmaddala}
\bibinfo{author}{Singh, S.}, \& \bibinfo{author}{Maddala, G.}
  (\bibinfo{year}{1976}).
\newblock \bibinfo{title}{A function for size distribution of incomes}.
\newblock {\it \bibinfo{journal}{Econometrica}\/},  {\it
  \bibinfo{volume}{44}\/}, \bibinfo{pages}{963--970}.
\bibitem[{Vuong(1989)}]{vuong1989likelihood}
\bibinfo{author}{Vuong, Q.~H.} (\bibinfo{year}{1989}).
\newblock \bibinfo{title}{Likelihood ratio tests for model selection and
  non-nested hypotheses}.
\newblock {\it \bibinfo{journal}{Econometrica}\/},  {\it
  \bibinfo{volume}{57}\/}, \bibinfo{pages}{307--333}.
\bibitem[{Waltman \& Van~Eck(2009)}]{waltman2009some}
\bibinfo{author}{Waltman, L.}, \& \bibinfo{author}{Van~Eck, N.~J.}
  (\bibinfo{year}{2009}).
\newblock \bibinfo{title}{{Some comments on Egghe's derivation of the impact
  factor distribution}}.
\newblock {\it \bibinfo{journal}{Journal of Informetrics}\/},  {\it
  \bibinfo{volume}{3}\/}, \bibinfo{pages}{363--366}.
\bibitem[{Wilfling \& Kr{\"a}mer(1993)}]{wilfling1993lorenz}
\bibinfo{author}{Wilfling, B.}, \& \bibinfo{author}{Kr{\"a}mer, W.}
  (\bibinfo{year}{1993}).
\newblock \bibinfo{title}{{The Lorenz-ordering of Singh-Maddala income
  distributions}}.
\newblock {\it \bibinfo{journal}{Economics Letters}\/},  {\it
  \bibinfo{volume}{43}\/}, \bibinfo{pages}{53--57}.

\end{thebibliography}







\end{document}